\documentclass{article}

\usepackage{PRIMEarxiv}

\usepackage[utf8]{inputenc} 
\usepackage[T1]{fontenc}    
\usepackage{hyperref}       
\usepackage{url}            
\usepackage{booktabs}       
\usepackage{amsfonts}       
\usepackage{nicefrac}       
\usepackage{microtype}      
\usepackage{lipsum}
\usepackage{fancyhdr}       
\usepackage{graphicx}       
\graphicspath{{media/}}     

\usepackage{amssymb}
\usepackage{pifont}
\usepackage{xcolor}
\definecolor{darkgreen}{HTML}{3C8031}
\newcommand{\greencheck}{{\color{darkgreen}\checkmark}}
\newcommand{\redxmark}{{\color{red}\ding{55}}}

\title{FEVA: Fast Event Video Annotation Tool
}

\author{
  Snehesh Shrestha, William Sentosatio, Huiashu Peng, Cornelia Ferm{\"u}ller, Yiannis Aloimonos\\
  The University of Maryland College Park\\
  College Park, MD 20740\\
  \texttt{snehesh@umd.edu}\\
}

\begin{document}
\maketitle


\begin{abstract}
Video Annotation is a crucial process in computer science and social science alike. Many video annotation tools (VATs) offer a wide range of features for making annotation possible. We conducted an extensive survey of over 59 VATs and interviewed interdisciplinary researchers to evaluate the usability of VATs. Our findings suggest that most current VATs have overwhelming user interfaces, poor interaction techniques, and difficult-to-understand features. These often lead to longer annotation time, label inconsistencies, and user fatigue. We introduce FEVA, a video annotation tool with streamlined interaction techniques and a dynamic interface that makes labeling tasks easy and fast. FEVA focuses on speed, accuracy, and simplicity to make annotation quick, consistent, and straightforward. For example, annotators can control the speed and direction of the video and mark the onset and the offset of a label in real time with single key presses. In our user study, FEVA users, on average, require 36\% less interaction than the most popular annotation tools (Advene, ANVIL, ELAN, VIA, and VIAN). The participants (N=32) rated FEVA as more intuitive and required less mental demand. The code and demo are available at \url{http://www.snehesh.com/feva}.
\end{abstract}

\begin{figure}[ht]
    \centering
    \includegraphics[width=\textwidth]{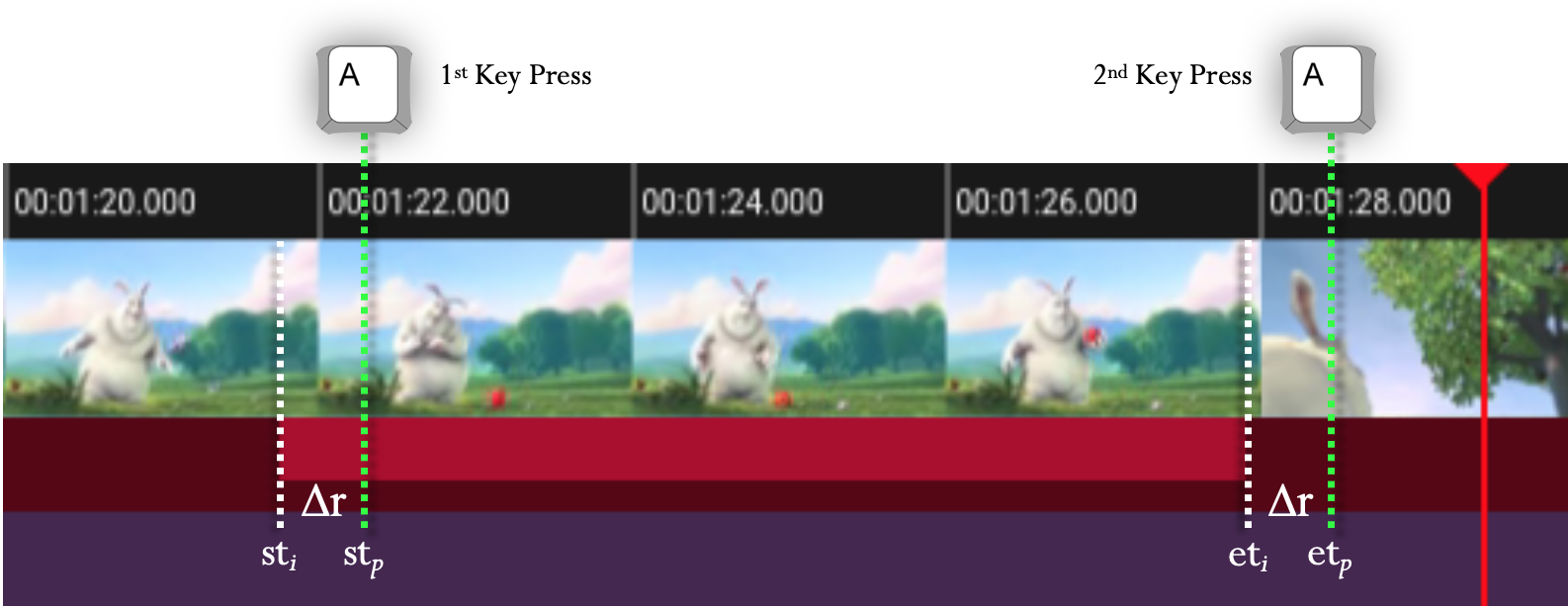}
    \caption{``Speed Label'' enables you to create annotations (red rectangle) in real-time by marking start (\(st_p\)) and end time (\(et_p\)) with a single key press respectively. FEVA automatically adjusts these times of the event annotation based on your reaction-time (\(\Delta r\)) in order to give you the most precise intended times (\(st_i\) and \(et_i\)).}
    \label{fig: feva_speedlabel_teaser}
\end{figure}

\keywords{Video Annotation Tools \and User Interface Design \and User Interaction} 

\section{Introduction}
Social scientists need to code conversations and behaviors of videotaped interviews and experiments that quickly add up to hours of footage \cite{broadhead2010playlearninfants, chen2012autismtherapy, Choi2022-it}. Computer scientists need datasets with appropriately labeled ground truth for machine learning that contains video clips spanning hundreds of hours \cite{Damen2020EpicDatasetCollection, Geiger2013KittiRawDataset, gemmeke2017audio}, or even thousands of hours \cite{abu2016youtube8m}. Annotating videos is thus time-consuming and tedious.

There are existing VATs \cite{29kipp2014anvil, 26aubert2012advene, 28wittenburg2006elan, hagedorn2008vcode, dutta2019via, 33halter2019vian} that offer a range of different features for video annotation activities. However, they often fail to meet the need of the researchers due to the steep learning curves with complicated features, overwhelming interfaces, and poor interaction techniques leading to longer annotation time, inconsistencies, and user fatigue.

For example, to analyze soccer games, researchers annotate player ball possessions, kicks, and assists. Annotating with tools that pause for the user to name the annotation every time, needing to annotate with mouse context menu options, and not allowing overlapping annotations make it an extremely tedious and time-consuming task. On the contrary, coding by hand is more straightforward. An annotator can play the video at a slower speed, use a clicker count or a stopwatch \cite{ward2020stopwatch} to lap the timestamps in real-time, then enter them into a spreadsheet when completed. As a result, many annotators still code by hand and rely on spreadsheets \cite{tannen2008researchertoolmismatch}. At scale, when you have hours of such footage, computer scientists often outsource or crowdsource such annotation tasks \cite{regneri2013actiondescriptionsvideos, vondrick2013crowdsourcedvideoannotation, escalante2016chalearn, riegler2016inflightdataset}. This can raise concerns about privacy and reliability \cite{lasecki2015tradeoffsprivacycrowdsourced}. To get around this, researchers blur faces to obfuscate participant identity. However, this could compromise the quality of the emotional judgment, causing inconsistencies in the results.

In this paper, we aim to design a video annotation tool that makes labeling tasks easy and fast. We interviewed researchers from different disciplines to understand standard practices, workflows, and tools used for video annotation. Specifically, we interviewed 13 researchers from 5 fields (neuroscience, behavioral psychology, film studies, and computer science). Researchers expressed reservations with existing VATs leading to avoiding them. We further surveyed 59 VATs (the list is available in the appendix \label{list_of_vat_appendix_reference}). We categorized their main features and interface design choices from firsthand experience and analyzed video tutorials for tools that are not accessible. According to the interviews and survey results, we propose five design criteria that would benefit video annotation activities, which are detailed in section \ref{sec: design_considerations}:

\begin{itemize}
    \item D1. Organize space based on operational workflow
    \item D2. Streamline high-frequency actions
    \item D3. Use algorithmic support when possible
    \item D4. Adopt what works and redesign what doesn't
    \item D5. Allow flexibility
\end{itemize}

\begin{figure}[h]
  \includegraphics[width=\textwidth]{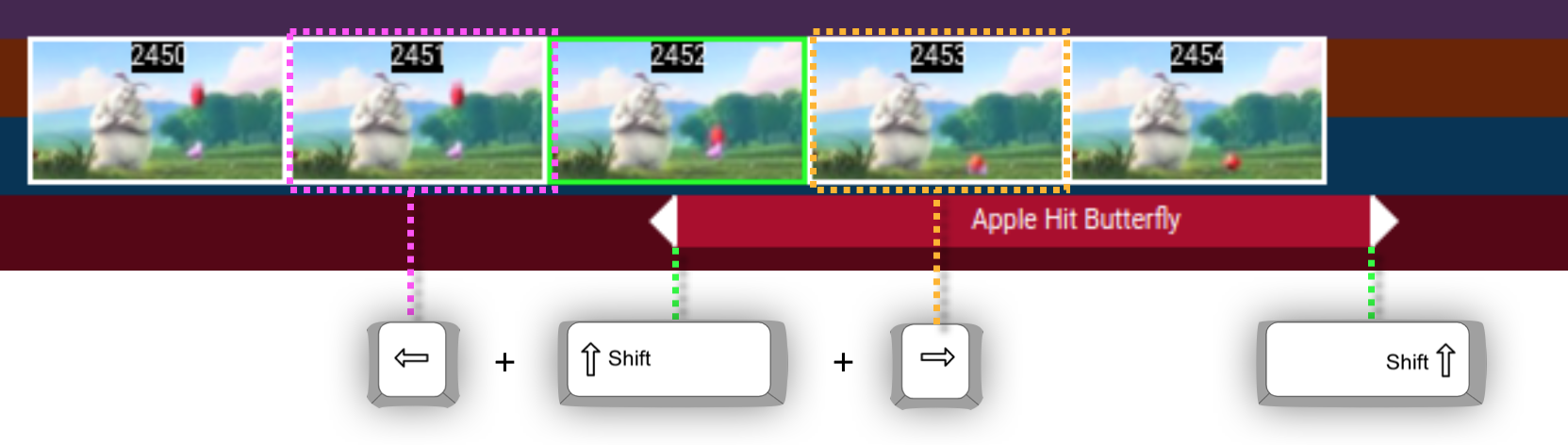}
  \caption{The shift keys activate the fine-tuning feature. The left and right shift keys correspond to the start and end time of the label, respectively. Pressing the arrow key while holding down the shift key adjusts the label's start or end time by a single frame.}
  \label{fig: feva_finetuning}
\end{figure}

These criteria inform the design of Fast Event Video Annotation (FEVA), a video annotation tool with streamlined interaction techniques and a dynamic interface that makes labeling tasks easy and fast. Simplified UI and features such as real-time labeling with reaction time adjustments (figure \ref{fig: feva_speedlabel_teaser}), and precise fine-tuning mechanisms (figure \ref{fig: feva_finetuning}), help annotators create a large number of accurate labels faster than any VAT.

To evaluate FEVA, we conducted two comparative studies. In the first study, we compared the number of inputs users need to complete a  task with FEVA versus five other event-based VATs \cite{29kipp2014anvil, 26aubert2012advene, 28wittenburg2006elan, dutta2019via, 33halter2019vian}. As seen in table \ref{tab: vat_benchmark}, on average, FEVA required 36.0\% fewer inputs than competing VATs to do the same task. In the second study, we asked 34 participants to perform annotations with one of the VATs in random order. We found that regardless of the background, users thought FEVA was more intuitive at 88\% and easier to use at 91\% of the time. Users also rated FEVA as requiring lower mental demand by 46\% (\textit{p}  < 0.00003), caused fewer frustrations by 62\% (\textit{p} <0.00147), had less physical demand by 41\% (\textit{p}  < 0.00187), and required less effort by 34\% (\textit{p}  < 0.00324).

In Summary, in this paper, our contributions are as follows:
\begin{itemize}
    \item a comprehensive understanding of \textbf{existing VATs} with interviews, tool surveys, and pilot tests.
    \item a list of \textbf{criteria} for VAT tool \textbf{design}
    \item \textbf{FEVA}, an event-based video annotation tool that lets you annotate faster and more accurately.
    \item a comparative study and a \textbf{user study} that demonstrates FEVA as more intuitive and requiring less effort while creating more consistent and accurate annotations.
\end{itemize}


\section{Related Work}
In this section, we introduce event-based annotation and tools we build upon, the workflow used in the annotation process, and finally, the user interface and the related interaction techniques. To understand the annotation workflow and goals, we interviewed researchers and surveyed the literature on the steps required to complete the desired annotations. We found two primary groups of annotators: ones that annotate with the assistance of VAT and ones that rely on more heuristic methods.

\subsection{Event Based Annotation}
Video annotation is the process of marking regions of interest (ROI) either a) spacial (object annotation, OA) or b) temporal (event annotation, EA). OA is marked on a single-time instance, referred to as frame-based annotation. EA is the period when the event occurs. OA is prevalent in the CV community for object detection and recognition, and bounding boxes \cite{park2019interactive, 22vondrick2013efficiently, 24bianco2015interactive, 25microsoft2019vott}, dots \cite{16biresaw2016vitbat, dutta2019via}, or polygon a.k.a segmentation \cite{mishra2009active, teo2015detection, barranco2018real} or masks \cite{16biresaw2016vitbat, 23russell2008labelme, 24bianco2015interactive} are used to mark the annotation. While some VATs also track objects \cite{fermuller1992tracking,mitrokhin2018event} over time \cite{dutta2019via, 16biresaw2016vitbat, 22vondrick2013efficiently, 25microsoft2019vott}, it is different from EA. EA focuses on what is happening in the scene, what the objects are doing, or what is done to the objects \cite{fermuller2018prediction,dessalene2021forecasting} rather than the objects themselves. So the start and the end time marks are used. EA includes behaviors, interactions, emotional response, speech, and movements \cite{dasiopoulou2011survey, 18burr2006vaca, 12brudy2018eagleview}. Their application range across disciplines from computer science for activity recognition \cite{Evanusa2022-oy,dutta2019via,yang2013cognitive,yang2015learning,teo2012towards,zampogiannis2015learning}, psychology for behavior analysis \cite{hagedorn2008vcode, satybaldiev2019coat, 18burr2006vaca}, to journalism to track and present stories over timelines \cite{bennett2020washingtonposttrumpchurchphotoop}. However, events annotation can be challenging due to the complex nature of temporal navigation, the ability to mark at the exact desired time of sliding events, use of available features of the tools to facilitate the annotation. Therefore, the focus of developing FEVA was to simplify the workflow and optimize the steps for the annotation of events to make annotation easier and faster.

\subsection{Heuristic Workflows}
Some researchers prefer not to use specialized video annotation tools. For example, some researchers thought it was easier to use a clicker to count the time certain events occurred. This is error-prone, less reliable, and makes it difficult to review the records in the future. Even though some VATs can provide a similar functionality \cite{dutta2019via} using keypresses, researchers showed hesitation using VATs with the fear of the initial setup time and the repeated learning curve for new coders. In another study \cite{ward2020stopwatch}, research assistants (RA) used a stopwatch to obtain the time taken "from when the OK button was pressed to when the device beeped to signal completion of the RR count." These clicker counts or stopwatch intervals are recorded either by hand with pen and paper or entered in a spreadsheet. In another workflow, an interpersonal relationship researcher described how, from a video of two people interacting, they asked RAs to respond to research questions set up by the researchers while watching a video. For example, multiple RAs focus on the entire interaction or a specific individual in the video and respond to a Likert scale on how attentive one of the partners was when the other spoke. These are either recorded on paper or online using Qualtrics or Google forms.

\subsection{Video Annotation Tools Workflows}
The primary workflow using a VAT does not differ from the heuristic methods for simple coding. The main difference is the higher learning curve and a longer setup process. However, as the coding gets more complex, heuristic methods have a diminishing return on speed as you have more annotators. The annotation process is much slower, needing to do a lot of steps typically tools might facilitate manually. However, not all VAT are created equal. The workflow design, screen layout of features, interaction steps, and techniques differ significantly across VATs.

Computer scientists and engineers design most VATs for computer vision, machine learning, and robotics research and applications \cite{dutta2019via, 18burr2006vaca, Damen2020EpicDatasetCollection, regneri2013actiondescriptionsvideos, abu2016youtube8m}. However, there are a handful of tools specifically designed for and used by other domains such as film studies \cite{33halter2019vian, hosack2010videoant} where VAT allows for character analysis and scene analysis. Journalists use VAT to annotate and synchronize events from multiple sources to present a cohesive story \cite{bennett2020washingtonposttrumpchurchphotoop}. In sports, teams annotate games and significant events and move to study for their teams \cite{sulser2014CrowdSport, 30piazentin2019historytracker}. While most tools have different focuses, the primary annotation is marking a binary exists or not and a temporal range of events of interest. To this end, events annotation is the foundation that makes further work easier or, in some cases, possible. So it is essential to have accurate annotations. Support for synchronized multi-camera views, micro and macro visualization of events in a timeline, and convenient searching and reviewing video features make FEVA a favorable choice in these areas.

\subsubsection{Layout design: Balance of features and space usage}
Some tools have a lot of features with complicated workflows and many permutations of features that can be customized and used, so the UI is packed with small windows, buttons, and layers of menu items to get to them. These tools take much longer to learn and get used to; however, they can be powerful once you learn them. Due to many features, such as media player controls, annotation controls, and visualizations, the available screen real estate can be quite challenging. So some tools group them into smaller windows \cite{33halter2019vian, 26aubert2012advene, 29kipp2014anvil}, organize them through multiple levels of menus \cite{26aubert2012advene, 28wittenburg2006elan}, or through different operational modes \cite{dutta2019via, 28wittenburg2006elan, 33halter2019vian}. While this helps organize the functionality, it isn't easy to access, and one needs a lot of practice to remember the various steps required to use the software. While there are other much simpler tools that serve a very niche purpose \cite{30piazentin2019historytracker, kim2013toolscape, hosack2010videoant, Levin2020timelinely, lagrue2019tdatontologydancevideos}, and the layout is simple, and they make excellent use of the screen real estate for the functionality they offer. These tools are easy to learn and start to use. However, the features are limited. Our interviews found that most researchers' and annotators' needs are in between. Most features are never used in the complex VAT, while simple VAT limits them from doing more than the niche they offer, and they need to pair with other tools to augment the gap to complete their needs. We designed FEVA with the motivation of creating a tool that is simple to navigate and use but comes with features that more complex tools offer without it being overwhelming that you have to take a course to use a tool.

\subsection{Adoption}
The keyboard shortcuts for VIA \cite{dutta2019via} are intuitive and practical, especially the hints that pop up based on the context is something most tools lack that FEVA adopts as well. FEVA uses popular shortcuts that have become the standard for media players, such as the spacebar to play or pause, ctrl+Z, and ctrl+Y to undo and redo, etc. VIAN \cite{33halter2019vian} is the only tool you don't have to select before dragging the label with the mouse, which is the same in FEVA. Advene \cite{26aubert2012advene}, ELAN \cite{28wittenburg2006elan} and VIA \cite{dutta2019via} provide alternative ways to visualize or execute the playback option, such as continuous mode. While useful in some instances, most annotators used the default way without changing, which could be confusing. While some tools \cite{dutta2019via} allows one to create a label when the movie is playing, it only expects the start time with a fixed length. Unfortunately, most users need to return to the label and readjust them. FEVA improves upon this interaction by allowing a second key press to mark the endpoint.

\section{Understanding the State-Of-The-Art VATs and Their Comparisons}

\subsection{Target Users}
To understand if researchers from different disciplines annotate their data, what that entails, and what the workflow looks like, we interviewed 3 neuroscience researchers, 3 behavior psychology researchers, 2 film study instructors, and 5 computer scientists. The interview was semi-structured to answer the following 3 questions:
\begin{itemize}
    \item \textbf{IQ1}: The nature of their research involving human studies, the kinds of data collected, and if it entails video.
    \item \textbf{IQ2}: The workflow in the data collection process, post-processing of these data, and code generation.
    \item \textbf{IQ3}: The structure and workflow for annotating the videos, and how the annotations are used after.
\end{itemize}

Post-interview, the responses were tallied and coded for technical challenges. The interview insights (II) are as follows:
\begin{itemize}
    \item \textbf{II1}. There were three primary temporal annotations.
1) A binary label to mark the presence or absence of certain events. For example, researchers were interested in counting "how many times a person touched their face as one indicator of how nervous the participants were."
2) A range label marks the beginning and end of an event of interest. "Participants annotate videos for specific moments. For instance, a couple might interact, then the researchers have each couple member watch the video and annotate whenever a specific thought or feeling occurred to them. They do this with both participants to see convergence and dissonance of thoughts, feelings, goals, etc., etc."
3) Certain kinds of labels, such as mood or scenario, lasted longer than an action label. "For instance, we might want to compute a rating of how responsive or caring one individual is to their relationship partner. So a Research Assistant might watch an entire interaction, focusing on a specific individual in the video, and then answer a question about how attentive they are when their partner speaks." Most VATs do a poor job of supporting these labels, so researchers had workarounds such as creating multiple label tracks, each dedicated to a specific response.
    \item \textbf{II2}: The time, effort, and cost for data annotation are exponentially high. So any system that made some improvements to make the annotation faster and more reliable was always a huge win. "We spend a lot of our RA hours doing these annotations. If there were a tool that could cut the time by even an hour, I definitely would be using that tool."
    \item \textbf{II3}: Even with very explicit codes, annotated data often had low temporal precision, so the agreement rules were relaxed. "Many RAs review and annotate the videos. There will be slight variations in when each RA thinks a certain event happened."
    \item \textbf{II4}: Collaboration and sharing of the annotation during the annotation process and analysis step was cumbersome and required multiple steps to be in sync between the teams. "...the students need to refresh the page if they are working on annotating the same clip simultaneously to see what their classmates are writing. That is also a problem for me since I like to add my comments while they are working (to encourage them to elaborate on points or explore a new point)."
    \item \textbf{II5}: Current VAT provided a poor interface for researchers to explore, analyze, and search the annotated data.
    \item \textbf{II6}: Crowd-sourced or AI-generated annotation often needed so much review that it was easier and faster for researchers' RAs to do the annotations. "So I used the [online automatic speech recognition tool]. With this, with the premium, I still have to go in and edit everything. The labels for the actions and the labels for the word-for-word speech need to match up in terms of start and end time. So creating the labels by hand is actually easier for me."
\end{itemize}

This paper focuses on II2, II3, and II5, which help shape the design decisions made in section \ref{sec: design_considerations}. 

\subsection{Comparing Video Annotation Tools specifications}
We created an extensive list of 59 VAT from the literature as listed in appendix \ref{appendix: list_of_VAT}. We were able to find their websites, download links or shared open source codes, or at the minimum online videos of either talk by the authors or how-to videos. We cataloged typical features most software supported and unique features and techniques distinctive to each tool. In this extensive survey, we share the table for the narrowed-down selected five tools. We detail the selection method in section \ref{section: SOTA_selection_method}. Two types of tables were created:
\begin{itemize}
    \item based on high-level taxonomy as shown in table \ref{tab: vat_taxonomy} and 
    \item based on features as shown in table \ref{tab: vat_features}
\end{itemize}

\begin{table}[!htp]\centering
\scriptsize
\begin{tabular}{lp{1.5cm}p{1.5cm}p{1.5cm}p{1.5cm}p{1.5cm}p{1.5cm}p{1.5cm}p{1.5cm}p{1.5cm}p{1.5cm}}\toprule
\textbf{SN} &\textbf{Features} &\textbf{Description} &\textbf{Advene} &\textbf{ANVIL} &\textbf{ELAN} &\textbf{VIA} &\textbf{VIAN} &\textbf{FEVA (Ours)} \\\cmidrule{1-9}
0 &Last Updated &Month Year &Jun 2020 &Mar 2019  &Mar 2020  &Jul 2020  &May 2020  &Sep 2020  \\\cmidrule{1-9}
1 &SW Platform &Cloud vs Edge &Edge &Edge &Edge &Edge &Edge &Cloud or Edge \\\cmidrule{3-9}
& &Native vs Web Based &Native &Native &Native &Web Based &Native &Web Based \\\cmidrule{3-9}
& &Modular vs Static &Static &Static &Static &Static &Static &Modular \\\cmidrule{1-9}
2 &License &Open Source vs Proprietary &Open Source &Proprietary &Open Source &Open Source &Open Source &Open Source \\\cmidrule{3-9}
& &Commercial vs Open Access &Open Access &Open Access &Open Access &Open Access &Open Access &Open Access \\\cmidrule{3-9}
& &Maintained vs Outdated &Maintained &Maintained &Maintained &Maintained &Maintained &Maintained \\\cmidrule{1-9}
3 &Cost &Free, low cost, vs Expensive &Free &Free &Free &Free &Free/Low Cost &Free \\\cmidrule{3-9}
& &One time vs Subscription &N/A &N/A &N/A &N/A &N/A &N/A \\\cmidrule{1-9}
4 &Collaboration &Single User &\greencheck &\greencheck &\greencheck &\greencheck &\greencheck &\greencheck \\\cmidrule{3-9}
& &Multi-User (Simultaneous) &\redxmark &\redxmark &\redxmark &\redxmark &\redxmark &\redxmark \\\cmidrule{3-9}
& &Crowd &\redxmark &\redxmark &\redxmark &\redxmark &\redxmark &\redxmark \\\cmidrule{1-9}
5 &Target Users &Technical vs Non-Technical &Technical &Technical &Technical &Technical &Technical &Both \\\cmidrule{3-9}
& &Academic vs Commercial &Academic &Academic &Academic &Academic &Academic &Academic \\\cmidrule{1-9}
6 &Input Type &Image &\redxmark &\redxmark &\redxmark &\greencheck &\redxmark &\redxmark \\\cmidrule{3-9}
& &Video &\greencheck &\greencheck &\greencheck &\greencheck &\greencheck &\greencheck \\\cmidrule{3-9}
& &Audio &\redxmark &\redxmark &\greencheck &\greencheck &\redxmark &\greencheck \\\cmidrule{1-9}
7 &Annotation Type &Object &\redxmark &\redxmark &\redxmark &\greencheck &\redxmark &\greencheck \\\cmidrule{3-9}
& &Action &\greencheck &\greencheck &\greencheck &\greencheck &\greencheck &\greencheck \\\cmidrule{3-9}
& &Events &\greencheck &\greencheck &\greencheck &\greencheck &\greencheck &\greencheck \\\cmidrule{3-9}
& &Hybrid &\redxmark &\redxmark &\redxmark &\greencheck &\redxmark &\greencheck \\\cmidrule{1-9}
8 &Annotation Approach &Manual &\greencheck &\greencheck &\greencheck &\greencheck &\greencheck &\greencheck \\\cmidrule{3-9}
& &Automatic &\redxmark &\redxmark &\redxmark &\redxmark &\redxmark &\redxmark \\\cmidrule{3-9}
& &Hybrid &\redxmark &\redxmark &\redxmark &\redxmark &\redxmark &\redxmark \\\cmidrule{1-9}
9 &Annotation Format &JSON &\redxmark &\greencheck &\redxmark &\greencheck &\redxmark &\greencheck \\\cmidrule{3-9}
& &XML &\greencheck &\redxmark &\greencheck &\redxmark &\redxmark &\redxmark \\\cmidrule{3-9}
& &SQL &\redxmark &\redxmark &\redxmark &\redxmark &\redxmark &\redxmark \\\cmidrule{3-9}
& &Others? &\greencheck &\greencheck &\greencheck &\greencheck &\greencheck &\redxmark \\\midrule
\bottomrule
\textbf{} &\textbf{} &\textbf{} &\textbf{} &\textbf{} &\textbf{} &\textbf{}
\end{tabular}
\caption{VAT taxonomy comparison table.}
\label{tab: vat_taxonomy}
\end{table}

\begin{table}[!htp]\centering
\scriptsize
\begin{tabular}{lllrrrrrrr}\toprule
\textbf{SN} &\textbf{Features} &\textbf{Description} &\textbf{Advene} &\textbf{ANVIL} &\textbf{ELAN} &\textbf{VIA} &\textbf{VIAN} &\textbf{FEVA} \\\cmidrule{1-9}
1 &Annotation types &Object Bounding Box &\redxmark &\redxmark &\redxmark &\greencheck &\redxmark &\greencheck \\\cmidrule{3-9}
& &Object Mask &\redxmark &\redxmark &\redxmark &\greencheck &\redxmark &\redxmark \\\cmidrule{3-9}
& &Object Dot &\redxmark &\redxmark &\redxmark &\greencheck &\redxmark &\greencheck \\\cmidrule{3-9}
& &Temporal Events &\greencheck &\greencheck &\greencheck &\greencheck &\greencheck &\greencheck \\\cmidrule{1-9}
2 &Playback controls &Play Pause FF RR &\greencheck &\greencheck &\greencheck &\greencheck &\greencheck &\greencheck \\\cmidrule{3-9}
& &Speed +/- &\greencheck &\greencheck &\greencheck &\greencheck &\greencheck &\greencheck \\\cmidrule{3-9}
& &Timeline Jump &\greencheck &\greencheck &\greencheck &\greencheck &\greencheck &\greencheck \\\cmidrule{1-9}
3 &Preview &Thumbnail Previews &\greencheck &\redxmark &\redxmark &\redxmark &\redxmark &\greencheck \\\cmidrule{1-9}
4 &Label &Multi-track &\greencheck &\greencheck &\greencheck &\greencheck &\greencheck &\greencheck \\\cmidrule{3-9}
& &Group tracks &\redxmark &\redxmark &\redxmark &\redxmark &\redxmark &\redxmark \\\cmidrule{3-9}
& &User-defined Label Types &\greencheck &\greencheck &\greencheck &\greencheck &\greencheck &\greencheck \\\cmidrule{3-9}
& &Show/Hide/Collapse/Expand &\redxmark &\redxmark &\redxmark &\redxmark &\redxmark &\redxmark \\\cmidrule{1-9}
5 &Speed Label &Sudo-Pedal &\redxmark &\redxmark &\redxmark &\greencheck &\redxmark &\greencheck \\\cmidrule{3-9}
& &Transcribing Pedal Support &\redxmark &\redxmark &\redxmark &\redxmark &\redxmark &\greencheck \\\cmidrule{1-9}
6 &Resize & &\greencheck &\greencheck &\greencheck &\greencheck &\greencheck &\greencheck \\\cmidrule{1-9}
7 &Move & &\greencheck &\greencheck &\greencheck &\greencheck &\greencheck &\greencheck \\\cmidrule{1-9}
8 &Add & &\greencheck &\greencheck &\greencheck &\greencheck &\greencheck &\greencheck \\\cmidrule{1-9}
9 &Delete & &\greencheck &\greencheck &\greencheck &\greencheck &\greencheck &\greencheck \\\cmidrule{1-9}
10 &Edit & &\greencheck &\greencheck &\greencheck &\greencheck &\greencheck &\greencheck \\\cmidrule{1-9}
11 &Import & &\greencheck &\greencheck &\greencheck &\redxmark &\greencheck &\greencheck \\\cmidrule{1-9}
12 &Import other formats & &\greencheck &\greencheck &\greencheck &\redxmark &\greencheck &\greencheck \\\cmidrule{1-9}
13 &Video Support &MP4 &\greencheck &\greencheck &\greencheck &\greencheck &\greencheck &\greencheck \\\cmidrule{3-9}
& &Others &\greencheck &\greencheck &\greencheck &\greencheck &\greencheck &\redxmark \\\cmidrule{1-9}
14 &Cameras &Multi-Cam &\greencheck &\redxmark &\redxmark &\redxmark &\redxmark &\greencheck \\\cmidrule{3-9}
& &Switch View &\greencheck &\redxmark &\redxmark &\redxmark &\redxmark &\greencheck \\\cmidrule{3-9}
& &Instant Switch View &\greencheck &\redxmark &\redxmark &\redxmark &\redxmark &\greencheck \\\cmidrule{1-9}
15 &History &Undo/ Redo &\greencheck &\redxmark &\greencheck &\redxmark &\greencheck &\greencheck \\\cmidrule{1-9}
16 &Search &Keyword &\greencheck &\redxmark &\greencheck &\redxmark &\greencheck &\greencheck \\\cmidrule{1-9}
& &Filter by label type &\greencheck &\redxmark &\greencheck &\redxmark &\greencheck &\greencheck \\\cmidrule{1-9}
17 &User Config &Remember/ Save &\redxmark &\redxmark &\greencheck &\redxmark &\redxmark &\greencheck \\\cmidrule{1-9}
18 &Modular/ API &Add-In Support &\greencheck &\redxmark &\greencheck &\redxmark &\greencheck &\redxmark \\\cmidrule{3-9}
& &Full Open Source Support &\greencheck &\redxmark &\greencheck &\greencheck &\redxmark &\greencheck \\\cmidrule{3-9}
& &Custom Layers &\redxmark &\redxmark &\redxmark &\greencheck &\greencheck &\redxmark \\\cmidrule{3-9}
& &Custom Tracks &\greencheck &\greencheck &\greencheck &\greencheck &\greencheck &\greencheck \\\cmidrule{1-9}
19 &Layers &Show/Hide Layers &\redxmark &\redxmark &\redxmark &\redxmark &\greencheck &\greencheck \\\cmidrule{3-9}
& &Human Joints Keypoints Support &\redxmark &\redxmark &\redxmark &\redxmark &\redxmark &\greencheck \\\cmidrule{3-9}
& &Human Bounding Box Support &\redxmark &\redxmark &\redxmark &\greencheck &\redxmark &\greencheck \\\cmidrule{3-9}
& &Human Mask Support &\redxmark &\redxmark &\redxmark &\greencheck &\redxmark &\redxmark \\\cmidrule{1-9}
20 &Export Support &Video Clips &\greencheck &\redxmark &\greencheck &\redxmark &\redxmark &\greencheck \\\cmidrule{3-9}
& &Image Frames &\redxmark &\redxmark &\greencheck &\redxmark &\redxmark &\greencheck \\\cmidrule{3-9}
& &Closed Caption &\greencheck &\redxmark &\greencheck &\redxmark &\redxmark &\greencheck \\\midrule
\bottomrule
\textbf{} &\textbf{} &\textbf{} &\textbf{} &\textbf{} &\textbf{} &\textbf{}
\end{tabular}
\caption{VAT features comparison table.}
\label{tab: vat_features}
\end{table}

\begin{figure}
  \includegraphics[width=\textwidth]{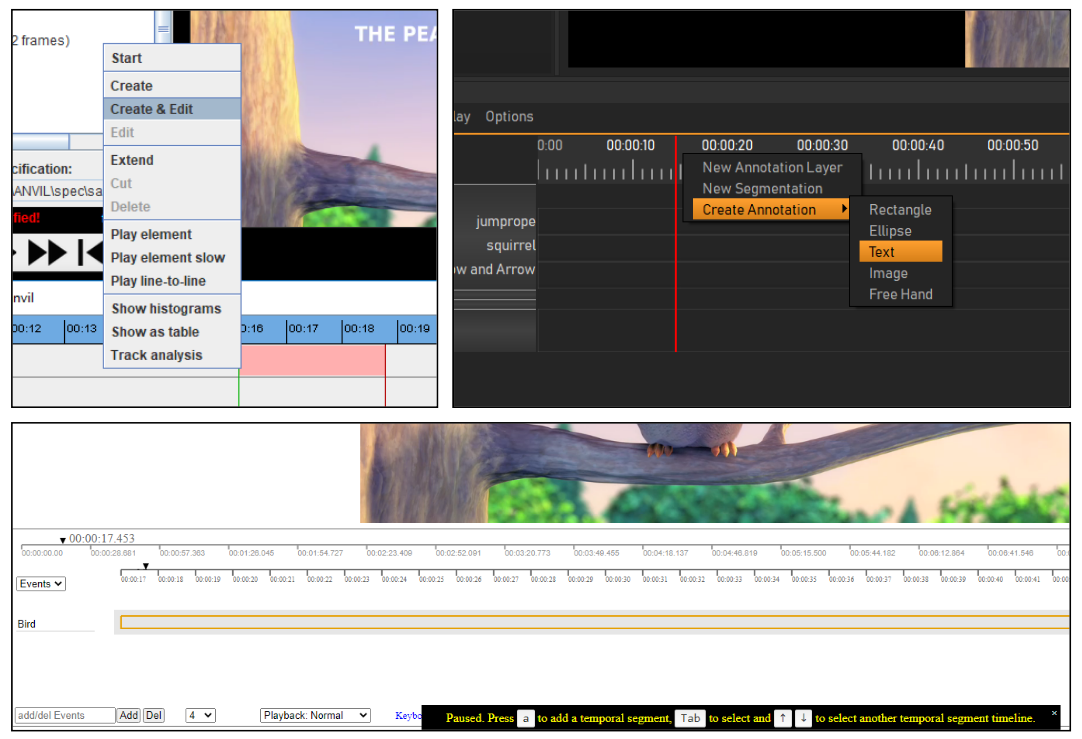}
  \caption{Example of steps needed to annotate in 3 different VATs. The Upper left screenshot is ANVIL, the upper right is VIAN, and the bottom screenshot is of VIA.}
  \label{fig: VAT_annotate_steps}
\end{figure}

As a contrasting difference, here is an example of the steps required to create a new annotation using 3 different VATs, as shown in figure \ref{fig: VAT_annotate_steps}. We make a more thorough step-by-step comparison in the evaluation section in \ref{section: interaction_benchmark}.
\begin{itemize}
    \item In the upper left corner is ANVIL. It requires you to double-click the starting point and click on the endpoint time mark to select the region. Then right-click the shaded area to select from the context menu to create and edit the label.
    \item On the upper right corner is VIAN. In VIAN, you right-click in the space between the tracks and the ruler. Then select create an annotation, then select text as the type for text annotation.
    \item In VIA, you create the track for the particular annotation and name it ("Bird" in this case), then play the video till the movie's starting point and hit the letter 'A' to create a fixed-size annotation. You can go back and adjust the annotation.
\end{itemize}

We surveyed the layout and the functionalities available in a number of events annotation software by using them to annotate videos or watch videos posted by the authors. We noticed the following:
\begin{itemize}
    \item Most VATs except for VIA \cite{dutta2019via} screen was filled with buttons, menus, and windows. While essential features were buried in multiple levels of menu items, there needed to be more screen real estate utilized. The new VAT requires to be simple to understand and easy to navigate.
    \item Observations through video instructions and pilot testing, we noticed that only limited features and controls were needed, primarily a) project and data management, b) annotation management, c) video navigation, d) annotations space, and e) the tool configurations. The new VAT needs a layout such that these features are upfront and eliminate unnecessary features or allocate them into rarely used spaces.
    \item The annotation visualization was limiting. An entire track was dedicated to a single label \cite{33halter2019vian}, and no overlapping time labels were possible \cite{29kipp2014anvil, 28wittenburg2006elan}. Only some \cite{dutta2019via, 26aubert2012advene} allow overlapping time labels. However, it is difficult to distinguish and manipulate the labels. The new VAT needs to organize the annotations automatically and better use annotation space.
    \item While some tools \cite{dutta2019via, hagedorn2008vcode} have good annotation UX controls to create and edit temporal placements, the use of mouse and keyboard had to be used interchangeably between annotation steps. Some tools \cite{29kipp2014anvil, 26aubert2012advene, 28wittenburg2006elan} have a very cumbersome way to move or resize annotations. Some tools provided a history option to undo/ redo \cite{26aubert2012advene, 28wittenburg2006elan}, while others provided no way to backtrack user mistakes or perform experimental steps. The new VAT needed simple mouse control and default keyboard shortcuts while allowing users to redefine them.
    \item All tools required you to stop and annotate except VIA \cite{dutta2019via}, which provided real-time annotation during video playback. However, VIA only allowed fixed-length annotation flags requiring adjustment of the endpoint later. Furthermore, VIA has poor visualization and lacks control for overlapping labels. The new VAT needs a fast real-time way to continue annotating without stopping.
\end{itemize}

\section{Design Considerations}
\label{sec: design_considerations}
According to the interviews, literature survey, and the use of the tools, we propose five design criteria that would benefit video annotation activities:

\begin{itemize}
    \item D1. \textit{Organize space based on operational workflow:} Features should be laid out in a logical flow of the workflow while not veering too far away from standard software conventions. Features not needed in that context should not be visible or active to free up valuable screen real estate. High-frequency controls and features should be in the middle of the screen.

    \item D2. \textit{Streamline high-frequency actions: } The highly repeated actions, such as creating annotations and fine-tuning them, should be optimized for easier and faster execution. Try to accomplish them with a single key press when possible. Stick to a single device (i.e., do not require some steps to use the keyboard and the other steps mouse movements, using up precious time in transitioning between the devices.) Leverage D1 when possible to minimize user interaction required to accomplish the task.
    
    \item D3. \textit{Use algorithmic support when possible: } Whenever possible, offload the user and rely on the algorithm to take on the burden. For example, when the timing is concerned, consider user reaction time and adjust for the lag in the user input from the intended time. Additionally, allowing external modules such as movement detection, human detection, speech detection, and recognition can offload users' need to find and annotate the events manually. However, we should be cautious that no matter how good machine learning modules are, no algorithm is 100\% accurate. These should be considered as only additional assistance and not to be completely relied on for ground truth generation.
    
    \item D4. \textit{Adopt what works and reinvent what doesn't: } Instead of reinventing the wheel, adopt from other tools what works well based on user tests or pilot tests and not instincts and personal preferences.
    
    \item D5. \textit{Allow flexibility: } While having D4, tested default input methods is great, adding flexibility by providing redundancies with mouse context menus and keyboard shortcuts might be more intuitive for some users. Additionally, let users redefine the key mapping as people have personal preferences.
\end{itemize}

\section{FEVA} 
\label{sec: feva_tool}

\begin{figure}[ht]
  \includegraphics[width=\textwidth]{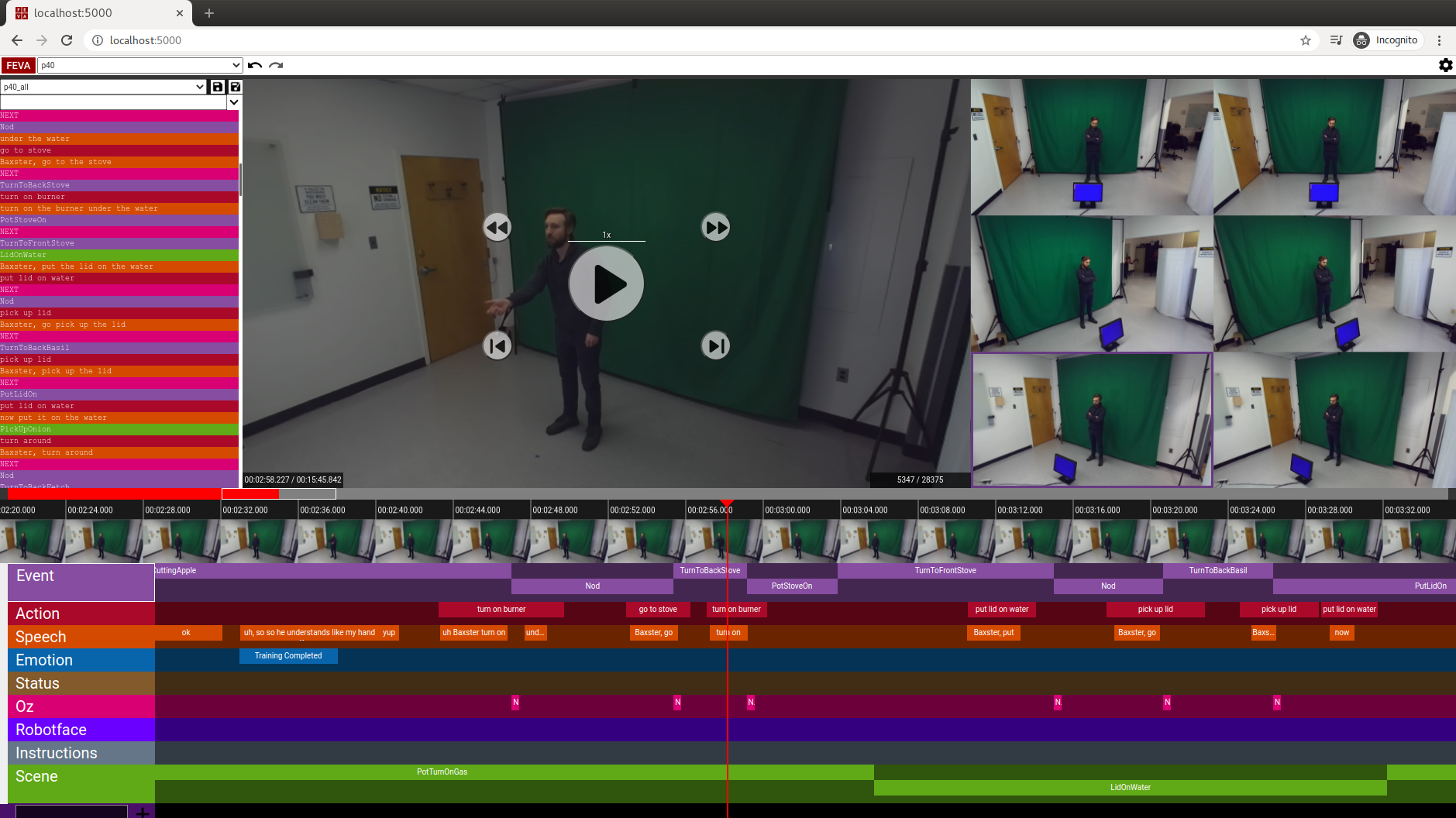}
  \caption{A screenshot of FEVA, where a participant interacting with a robot, is being annotated.}
  \label{fig: feva_layout_screenshot}
\end{figure}

The design considerations inform the minimalist design of FEVA, with most of the screen real estate allocated for the video and the annotation. 
Inspired by the clicker/ stopwatch methods and transcribing pedal, we created the speed label. This feature lets you create labels during video play, where a user can press a key to mark the starting and stopping points and continue to watch, creating multiple labels with desired lengths. The system additionally accounts for the user's reaction time and adjusts the marked times as seen in figure \ref{fig: feva_speedlabel_teaser}.

The speed label enables users to annotate in real-time, the fine-tuning control lets the user make frame level adjustments, and the label organizer arranges the label without ever overlapping them for direct access. The flat UI is responsive and aesthetically pleasing. Using F-shape design \cite{neilson2006fshapedpattern}, the initial workflow items are on the top left, while the most viewed items are in the middle of the screen, and the most interacted components are at the bottom of the screen. Keeping most used elements in the hot zones while less used elements such as camera selection and configuration buttons are on the right-hand side in the less noticed area, making them easily tuned out unless needed. The UI has redundant user input for flexible and fast interaction to cater to novice and expert users. And underlying context and configuration-dependent UI model comes to life only when you need them. Expert users can annotate videos faster than in real-time by using media speed control and speed labeler without touching the mouse. Researchers can also use the zoom feature to visualize and analyze labels at a micro or macro level. FEVA UI can display the most number of labels on one screen without losing their meaning.

\subsection{Workflow}
To create any project, upon selecting a video, FEVA imports it and creates an empty annotation file, also referred to as a label or the dataset file (D1, D2). Default tracks will be loaded, which can be customized from the configuration (D3). Users can add or remove tracks. To play or pause the movie, users can use the media player overlay on the main video or the keyboard shortcut 'spacebar.' You can use your mouse scroll button to navigate the video in the timeline. You can also use the arrow keys with or without the 'ctrl' key to move the video by different amounts. You can click and drag the filmstrip, roll over the global timeline and click at the desired time, or double-click the filmstrip or the local timeline window (D1, D2, D5). You can also change the movie playback speed. You can zoom in and out at different time intervals using the + and - icons around the white box on the global timeline or roll over the timeline and ctrl and scroll.

To annotate, users can right-click on the tracks and select the label type they want to create. Users can also hit the letter 'A' key on the keyboard to mark the start and a second time to mark the stop. See figure \ref{fig: feva_speedlabel_teaser}. This is called the speed label, as you can keep annotating without stopping the movie. Speed annotation requires a two-pass, but in our pilot tests, the speed label is at least 1.5x faster than the traditional methods.

Following left-to-right and top-to-bottom conventions, workflow such as loading the project, dataset, and manipulating the video and labels are organized in that order. The main video is at the center of the screen occupying the most space. Below the video are the annotation tracks that follow conventions and eyes and the hand layout of users' gaze and action areas. See figure \ref{fig: feva_layout} and figure \ref{fig: feva_layout_screenshot}.

\subsection{View/ Layout}
\begin{figure}
  \includegraphics[width=\textwidth]{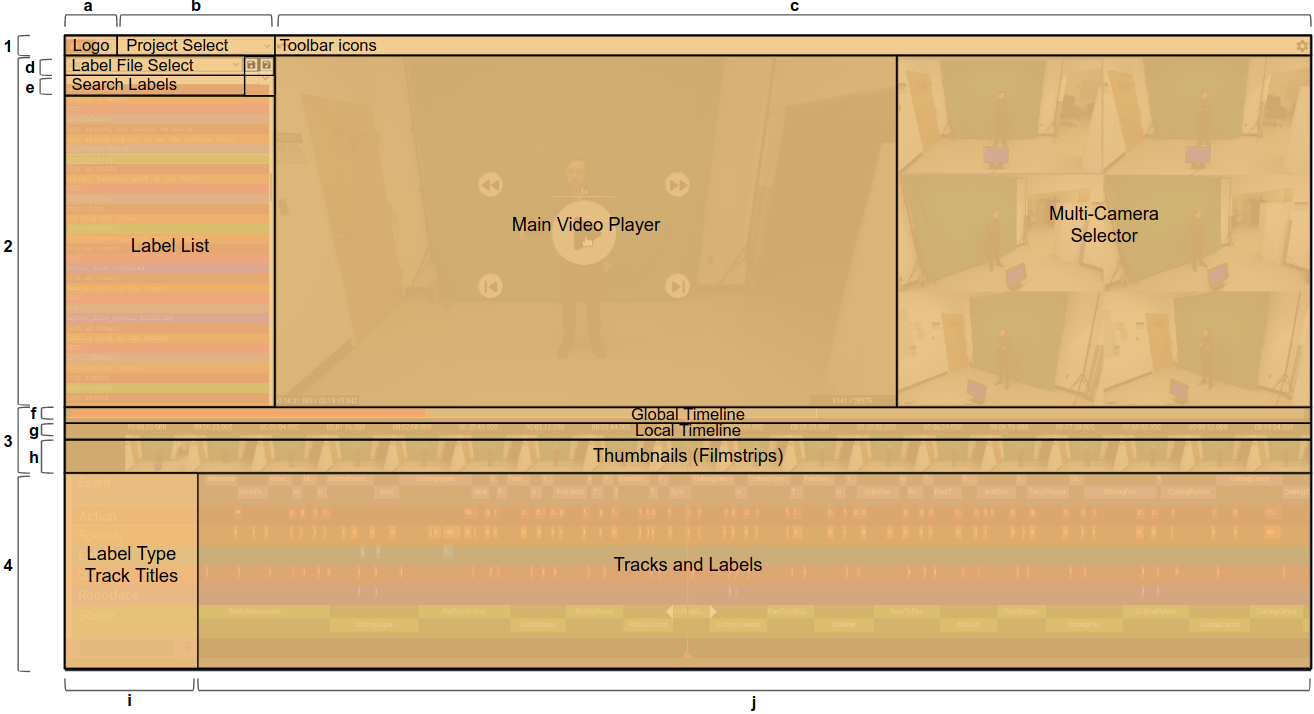}
  \caption{FEVA Screen Default Layout black boxed areas are 1) Toolbar, 2) Label list, main video player, and multi-camera selector, 3) Video navigation timeline, and 4) Label tracks. Key components are a) Logo, b) Project selection, c) toolbar icons, d) Label data file selection, e) Search bar, f) Global progress bar timeline, g) Local timeline ruler, h) Filmstrips, i) Label type, and j) tracks and labels.}
  \label{fig: feva_layout}
\end{figure}

As seen in figure \ref{fig: feva_layout}, 'b' shows the project selector from which you create a new project and import videos. The 'd' shows your label file selector to create, load, save, merge, import, and export labels. You can search using the 'e' and filter labels by type. You can double-click the label from the '2' label list to find the corresponding label in the timeline 'j.' You can see the thumbnail preview in 'h' along with the current time window 'g' and the global timeline 'f.'

\subsection{Control: User input system}
Every media control controlled by the mouse also has its associated keyboard shortcuts. For flexibility and efficiency, GUI for novice users that are based on principles of recognition rather than recall \cite{neilson2006usabilityheuristics}, and shortcuts for expert users for faster control. Users can press the play button overlayed on the video or use the spacebar key at any point to play or pause the video. You can also choose to speed up or slow down. To create an annotation while the video is in play, you can press the letter 'A' twice to indicate the start of the label and end the label. This can be customized from the configuration. A blank label is created after adjusting for your reaction time. You can also double-click an empty space on a track or right-click the tracks and select a label type you wish to create. While these shortcuts were selected to be consistent with existing standards from other VATs, all shortcuts can be user-defined easily in the user configuration.

To navigate, you can use your mouse to scroll or click and drag the filmstrips, double click the desired point in the filmstrip or the global progress bar. You can also jump to a specific label from the label list by double-clicking it from the label list. Users can choose what they prefer by providing multiple redundancies with both keyboard and mouse and keyboard shortcuts that can be re-customized by the user.

\subsection{Model/ State: Underlying UI support system}
The UI makes use of the limited screen space (x-y plain) by using the z-axis to layer components displayed and state changes based on contexts such as if the video is playing, if labels are selected, if labels are being edited, or if your mouse is hovering over a component or a specific feature is enabled in the configuration. Every area is compacted with features that feel intuitive based on affordance users naturally would assign those components. An example is the video player area. When the video is playing, one would only see the video. When a mouse pointer hovers over, media controls, current time and frame number, and layer control are displayed. From the configuration, you can also enable showing or hiding the video, human body keypoints \cite{cao2019openpose}, the bounding box of humans or objects, segmentation masks, etc., that are extensible for researchers to customize.

\section{Evaluation}
To evaluate FEVA, we compared FEVA with existing state-of-the-art (SOTA) VAT with two studies.
\begin{itemize}
    \item Interaction Benchmark: To evaluate the theoretical limits of how fast users could annotate with each VAT, we counted the number of user inputs required to perform various tasks.
    \item User Study: To evaluate user experience based on the user's perceived workload with each VAT, we conducted user studies where the users provided feedback based on their experience.
\end{itemize}

\subsection{The State-of-the-Art VAT Selection Method}
\label{section: SOTA_selection_method}
We first created a master list of highly cited VAT that we could download and use. In this list, we only included software that supported temporal annotation that could be downloaded, installed, and run without taking extreme measures for practical reasons. Therefore, VCode \cite{hagedorn2008vcode}, SVAT \cite{32rehatschek2007svat}, and VACA \cite{18burr2006vaca} could not be included. Tools that were too specific such as ToolScape \cite{kim2013toolscape}, HistoryTracker \cite{30piazentin2019historytracker}, and CASAM \cite{21hendley2014casam} due to missing functionalities such as start and end time, were removed from the list. Tools focused on crowd-sourcing such as Glance \cite{20lasecki2014glance} and CoAT \cite{satybaldiev2019coat}, were excluded as we conducted a single-user study. Using these criteria, we could narrow down the tools to be compared in the study. EagleView \cite{12brudy2018eagleview} was extremely unstable to run, so we could not test them. We narrowed down to Advene \cite{26aubert2012advene}, ANVIL \cite{29kipp2014anvil}, Elan \cite{28wittenburg2006elan}, VIA \cite{dutta2019via}, and VIAN \cite{33halter2019vian}.

\subsection{Interaction Benchmark}
\label{section: interaction_benchmark}
To compare the steps required to do a particular task with each VAT, we counted the number of clicks, double-clicks, mouse movements, and keyboard key presses and took a cumulative sum as seen in table \ref{tab: vat_benchmark}. If there were multiple ways of completing a task, we included the fastest method for that tool. For example, if you can press Ctrl+N to create a new project (keypresses count = 2) or can move your mouse to the main file menu, click the file, move the mouse to a new project, and click on the menu item (mouse move = 2 and mouse clicks = 2, total = 4), then we took the lesser of the two.

Table \ref{tab: vat_benchmark} shows the 15 tasks considered for the evaluation. These included basic setting, label creation, and manipulation tasks. We selected tasks that the majority of the VATs could do. If a VAT missed a feature, we assigned the worst count received by competing with the VATs. For example, \cite{dutta2019via} does not support "undo" or "redo" and received a count of 2.

The number of inputs required in FEVA is significantly less than the SOTA, as shown in table \ref{tab: vat_benchmark}. On average, FEVA requires 36\% less input than the SOTA. Based on the T-test, FEVA required significantly less input than all tools except VIA, which was not statistically significant.

\begin{table}[!htp]\centering
\scriptsize
\begin{tabular}{llccccccc}\toprule
\textbf{SN} &\textbf{Tasks} &\textbf{Advene} &\textbf{ANVIL} &\textbf{ELAN} &\textbf{VIA} &\textbf{VIAN} &\textbf{FEVA} \\\cmidrule{1-8}
1 &Create a project + Import a video &7 &8 &8 &7 &14 &\textbf{3} \\\cmidrule{1-8}
2 &Create a single label &4 &6 &7 &\textbf{2} &5 &\textbf{2} \\\cmidrule{1-8}
3 &Create multiple labels &8 &12 &14 &\textbf{3} &9 &\textbf{3} \\\cmidrule{1-8}
4 &Create and name label &\textbf{4} &9 &7 &7 &9 &7 \\\cmidrule{1-8}
5 &Edit labels &4 &7 &\textbf{3} &\textbf{3} &\textbf{3} &5 \\\cmidrule{1-8}
6 &Resize labels &6 &5 &6 &\textbf{4} &\textbf{4} &\textbf{4} \\\cmidrule{1-8}
7 &Move labels &6 &12 &6 &5 &\textbf{4} &\textbf{4} \\\cmidrule{1-8}
8 &Change label type &6 &6 &6 &6 &6 &\textbf{4} \\\cmidrule{1-8}
9 &Delete labels &\textbf{3} &5 &4 &\textbf{3} &\textbf{3} &\textbf{3} \\\cmidrule{1-8}
10 &Find labels &3 &3 &5 &5 &3 &\textbf{2} \\\cmidrule{1-8}
11 &Save labels &\textbf{2} &\textbf{2} &\textbf{2} &\textbf{2} &\textbf{2} &\textbf{2} \\\cmidrule{1-8}
12 &Load labels &\textbf{4} &\textbf{4} &\textbf{4} &\textbf{4} &8 &\textbf{4} \\\cmidrule{1-8}
13 &Navigate video &2 &6 &2 &\textbf{1} &2 &\textbf{1} \\\cmidrule{1-8}
14 &Play/ Pause video &2 &\textbf{1} &2 &\textbf{1} &\textbf{1} &\textbf{1} \\\cmidrule{2-8}
&Play only label video &6 &5 &4 &3 &6 &\textbf{2} \\\cmidrule{1-8}
15 &Undo/ Redo &\textbf{2} &2 &\textbf{2} &2 &\textbf{2} &\textbf{2} \\\cmidrule{1-8}
\textbf{} &\textbf{TOTAL SCORE} &69 &93 &82 &58 &81 &\textbf{49} \\\midrule
\textbf{} &\textbf{FEVA Faster by} &\textbf{29\%} &\textbf{47\%} &\textbf{40\%} &\textbf{16\%} &\textbf{40\%} \\\midrule
\textbf{} &\textbf{p-value} &0.0255 &0.0013 &0.0149 &0.1321 &0.0223 \\\midrule\bottomrule
\textbf{} &\textbf{} &\textbf{} &\textbf{} &\textbf{} &\textbf{} &\textbf{}
\end{tabular}
\caption{The list of tasks done using the fastest possible methods in each software (shortcuts where applicable). Each number reflects a cumulative sum of mouse clicks, double clicks, movement, and key presses. The last row shows how much FEVA is faster than the SOTA in percent (\%) and the T-test p-values.}
\label{tab: vat_benchmark}
\end{table}

\subsection{User Study}
\label{section: user_study}
To evaluate user experience, we conducted a user study where participants used two VAT, FEVA, and one another selected SOTA VATs (Advene, ANVIL, ELAN, VIA, and VIAN) in a round-robin fashion. We counterbalanced the order of the two tools by alternating the order with the next participant. Due to the COVID-19 regulations, we conducted our study via Zoom remote shared screen and control feature. This introduced some lag in the user experience. However, since both tools were remote, we assumed that the effects of the lag on the outcome were not significantly discriminant. Participants were given an approximate time range where an event occurs with clear descriptions of the events to annotate, for instance, as seen in figure \ref{fig: feva_task_example}, "between 4 minutes and 20 seconds and 4 minutes 40 seconds, please annotate bunny jump roping." Participants completed approximately 24 tasks until they gave up on the tool. After completing all the tasks in the first software, they filled out the NASA Task Load Index \cite{hart2006nasa} questionnaire with a 5-point Likert scale. And this was repeated with the second tool.

\subsubsection{Participants}
We recruited 34 participants in the University community via email and social media forums. In our study of N=32, the participants were 53\% male and 47\% female, had a mean age of 30.4 +/- 5.9, with 84\% not having video annotation experience, and 66\% had no experience with video editing. Two participants had to be dropped due to zoom connection issues and are not counted in N=32.

\subsubsection{Procedure}
\begin{figure}
  \includegraphics[width=\textwidth]{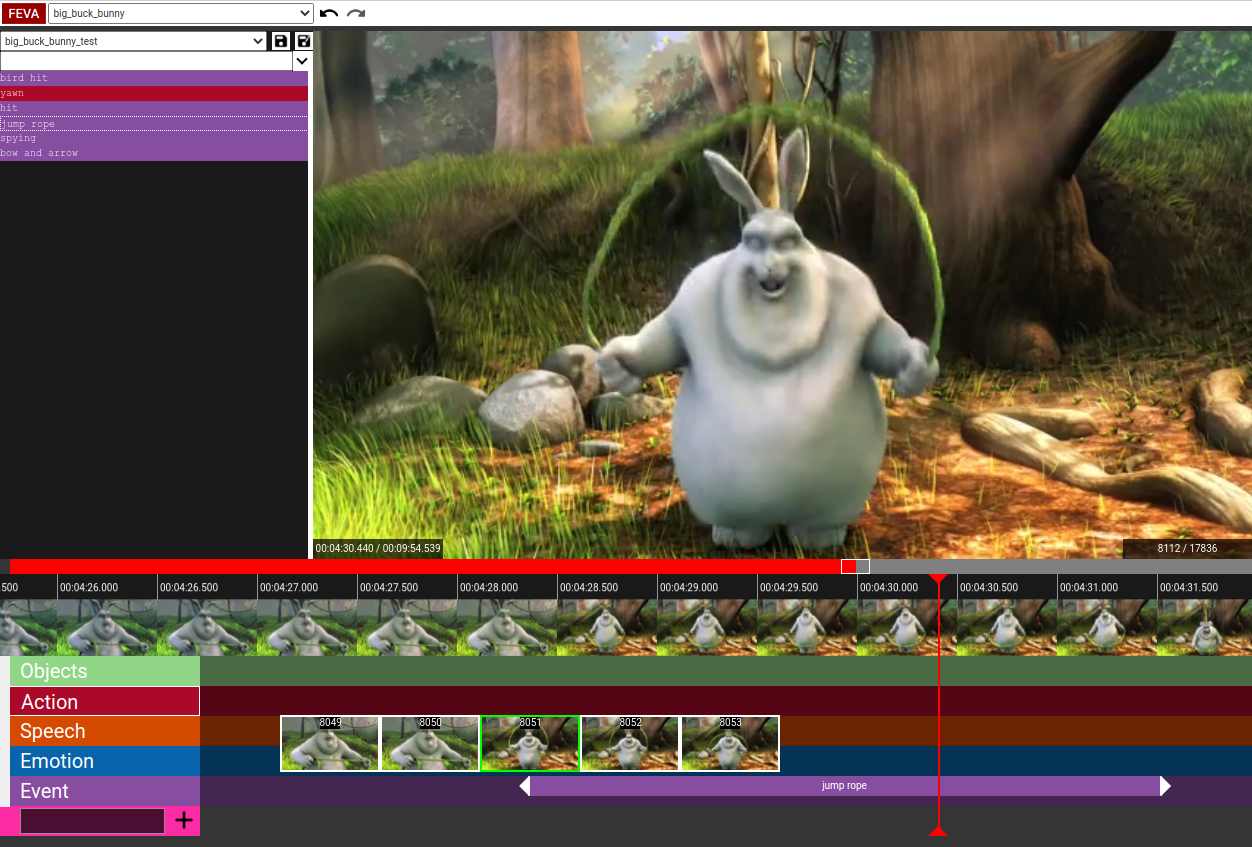}
  \caption{An example of a task where a participant was asked to annotate an event of the bunny jump roping between 04:28 minutes mark and 04:31 minutes mark in FEVA.}
  \label{fig: feva_task_example}
\end{figure}

For this study, we trained annotators for 3 minutes by watching a short training video that taught the basics of media controls, video timeline navigation, and how to create and edit annotations, followed by practice trials for each item with the research coordinator answering any questions. They then spent another 2 minutes exploring the tool on their own. The participants spent the next 5 minutes practicing the tasks assigned individually by the researcher where they were allowed to ask questions. Once they got comfortable, they were randomly given four categories of tasks shown in the list below, with each type of task repeated at least three times. The tasks chosen were the most fundamental and repeated tasks annotators must do during video annotation. The tasks ranged in complexity, with some tasks requiring combinations of the fundamental steps. For instance, some tasks simply asked the participant to navigate the video to 2 minutes and 20 seconds.
In contrast, others asked participants to annotate three consecutive events between a specific time and name them appropriately. They were no time limits to perform the tasks. They worked on the standard freely available "Big Buck Bunny" video which is approximately 10 minutes long at 720p resolution. Figure \ref{fig: feva_task_example} demonstrates one example task. After completing all the tasks with the first tool, the participants filled out the NASA TLX workload questionnaire and repeated the tasks with the second VAT.

We conducted the following four categories of basic tasks during the user study:
\begin{itemize}
    \item Navigate the video a) play/ pause the video, b) jump to a specific point in the timeline, and c) jump to a precise point where a particular label is.
    \item Label Creation a) Create a new label at a specific time with a specific length,  b) Create a label when a participant shows a specific behavior (e.g., a character yawns, eats an apple, etc.), and c) create multiple labels in a row.
    \item Label Content Manipulation a) Write text annotation for a label created and b) Modify annotation text.
    \item Label Temporal Manipulation a) Move the label by a specific number of seconds and b) Resize the label to change its starting time or ending time to match a specific behavior by the person in the video
\end{itemize}

\subsubsection{Results}
\label{section: results}
In this study, on average, the users felt less metal demand by \textbf{46\%} (\textit{p}  < 0.00003) with FEVA than the SOTA, less physical demand by \textbf{41\%} (\textit{p}  < 0.00187), less effort was required by \textbf{34\%} (\textit{p}  < 0.00324), and felt less frustration by \textbf{62\%} (\textit{p} <0.00147). The difference in the temporal demand and the performance level indexes was not significant. We attributed this to there not being a time limit enforced during the study, and except for one user on VIAN, where the user gave up, all other users completed all the tasks.

\begin{table}[!htbp]\centering
\scriptsize
\begin{tabular}{p{2cm}p{1cm}p{1cm}p{1cm}p{1cm}p{1cm}p{1cm}p{1cm}}\toprule
\textbf{FEVA} &\textbf{Mental Demand} &\textbf{Physical Demand} &\textbf{Temporal Demand} &\textbf{Performance Level} &\textbf{Effort} &\textbf{Frustration Level} \\\cmidrule{1-7}
\textbf{MEAN (n=32)} &1.8 &1.5 &1.8 &4.3 &2.3 &1.7 \\\cmidrule{1-7}
\bottomrule
\textbf{} &\textbf{} &\textbf{} &\textbf{} &\textbf{} &\textbf{} 
\end{tabular}
\caption{Shows FEVA's average score on a NASA Task Load Index with a 5-point Likert scale.}

\begin{tabular}{p{2cm}p{1cm}p{1cm}p{1cm}p{1cm}p{1cm}p{1cm}p{1cm}}\toprule
\textbf{FEVA vs. (\%)} &\textbf{Mental Demand} &\textbf{Physical Demand} &\textbf{Temporal Demand} &\textbf{Performance Level} &\textbf{Effort} &\textbf{Frustration Level} \\\cmidrule{1-7}
\textbf{Advene (n=5)} &75\% &57\% &0\% &14\% &67\% &183\% \\\cmidrule{1-7}
\textbf{ANVIL (n=8)} &25\% &36\% &-8\% &3\% &17\% &6\% \\\cmidrule{1-7}
\textbf{ELAN (n=8)} &29\% &42\% &13\% &12\% &38\% &83\% \\\cmidrule{1-7}
\textbf{VAI (n=8)} &44\% &36\% &6\% &11\% &22\% &36\% \\\cmidrule{1-7}
\textbf{VIAN (n=3)} &120\% &40\% &20\% &17\% &83\% &140\% \\\cmidrule{1-7}
\textbf{MEAN (n=32)} &\textbf{46\%} &\textbf{41\%} &5\% &10\% &\textbf{34\%} &\textbf{62\%} \\\cmidrule{1-7}
\bottomrule
\textbf{} &\textbf{} &\textbf{} &\textbf{} &\textbf{} &\textbf{} 
\end{tabular}
\caption{Shows how FEVA compared to the other VAT. A positive number indicates how much people perceived FEVA to be better than other VATs in the NASA TLX respective six dimensions, and a negative number indicates how much worse.}\label{tab: user_study}
\end{table}

\subsection{User Feedback}
\subsubsection{The Good}
On average, users expressed FEVA was more intuitive 88\% of the time and that FEVA was easier to use 91\% of the time. The features users liked the most were the speed label, fine adjustments, "cooler feel," locating label and label playback. A few users wished they could go back and change their feedback for the first tool once they used the second tool. This was typical when they felt they gave the first tool too high scores after using FEVA. In contrast, this did not happen when it was the other way around. One user said the user wanted to start a fan club and wanted to volunteer to annotate because it was "so fun."

\subsubsection{The bad}
One user mentioned that the user preferred traditional windowed UI for serious work. So the user thought FEVA looked too mainstream tablet app-like. Another user stated, "while I think it was fine for me, I don't think my mom will be able to use either of the tools. So I gave low scores to both of them." A few users complained that they did not like pressing the enter key to confirm the label after editing them. Clicking elsewhere,  causing the loss of what was just typed as a cancel feature, was not popular.

\subsubsection{The ugly}
The majority of the confusion, however, was about the global and the local timeline due to needing a clear separation and sharing the same preview component. Participants remotely controlling the UI of the VATs over a Zoom call on the research coordinator's computer noticed a lag in the effect of their actions. Some users complained that the UI did not update fast enough due to the Zoom lags. "Maybe because I am controlling your computer through Zoom, but a huge delay made it harder for me to resize the labels."

\section{Implementation}

\subsection{Framework and Dependencies}
We used ReactJS \cite{react2013}, an efficient component-based JavaScript library, and wrote the architecture to be lightweight and responsive, so it works on most people’s computers. The installation has only two dependencies of Python and Flask. The front end relies on standard HTML, Javascript, ReactJS, and CSS. We designed all the controls to optimize for performance and flexibility to customize. We detail the UI layout breakdown in figure \ref{fig: feva_layout} and section \ref{sec: feva_tool}.

\subsection{Architecture}
FEVA uses a simple server-client architecture typical for many web-based applications. The server side runs on Python 3.5x or newer with Flask as the webserver. The server side primarily handles servicing data (web content, annotation data, and video streaming) when requested by the client-side application. For FEVA, most of the modules and the design are on the client side. We show more details of all the modules and their interaction with other modules in the block diagram in appendix \ref{appendix: feva_client_architecture}.

\begin{figure}[ht]
  \includegraphics[width=\textwidth]{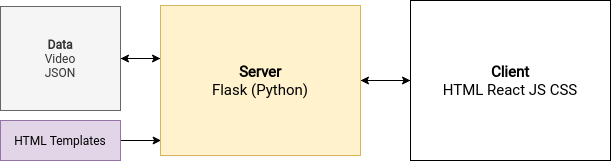}
  \caption{FEVA Client-Server Architecture Diagram. We include a more detailed block diagram of the different modules on the client side in the appendix \ref{fig: feva_architecture_client}.}
  \label{fig: feva_architecture}
\end{figure}


\section{Discussion}
This is the first version release of our tool FEVA, where we focused on building the fundamental tool while streamlining the user interface and interactions to make annotating events faster, intuitive, robust, and more accurate. While these early results look promising, there are more research questions that need to be further explored.

\subsection{User Study}
In our pilot and user study, we conducted a short-duration study. In our user study, we assumed that 15 minutes was sufficient time for participants to learn and practice annotating, which is how we designed the first evaluation of comparing multiple VATs. However, we need to further our research by conducting a longitudinal user study to understand the impact of our design on users as they get more comfortable with the software.

\subsection{User Input}
In the input sector, we considered the keyboard and mouse/ trackpad as the interaction devices at this stage. Still, we need to expand this to other kinds of inputs, such as touch, speech, and gesture, to explore the potential benefits of multi-modal methods.

\subsection{Target Users}
As more general public gets involved in the coding process, we focused this study so anyone can participate in the coding process. Future studies to gather feedback from seasoned coders will be valuable in understanding how they use VAT.

\subsection{VAT benchmarking study}
In section \ref{tab: vat_benchmark} study, we counted the number of inputs as the pilot study showed the correlation between the number of clicks and the time taken to complete a task. A more comprehensive study should be considered, including the mouse movements and time taken that can reflect user confusion and a more accurate performance metric for completing the task.

\subsection{Implicit}
In section \ref{section: user_study}, we focused on the user's perception to reflect their experience. Future studies should expand to more quantitative measures implicit in evaluating user confusion, performance, and success. We would also like to conduct more studies to understand labeling consistency.

\subsection{The layout}
Based on user feedback, lessons learned from our observations, and the process of comparing with existing tools, we could have done better. The multi-camera selector layout takes up a lot of screen real estate. Many users found the global timeline being so close to the local timeline and the thumbnail view without any separation confusing and more challenging to get used to. We have planned to redesign those experiences in the subsequent versions. We will focus on several optimization opportunities in the next release to make FEVA even faster.

\subsection{Future Work}
Beyond the incremental improvements, there were key features that we have planned for the future:
\begin{itemize}
    \item Adaptive: All the input controls were linear. We plan to explore dynamic and adaptive control systems to various new interaction techniques for faster annotation and a more intuitive experience.
    \item Extensible: An easier workflow for the open-source community to extend the features.
    \item AI assisted: While this had some algorithmic support, better integration with machine learning and deep neural network models is needed. We will further research how AI can augment the annotation process while exploring ways to inform the users of these models' inaccuracies, uncertainties, and inherited bias.
    \item Remote videos: While our internal prototypes support YouTube, there are optimization opportunities that we need to explore before they can be used seamlessly as an alternative to the local MPEG videos. We will also explore other online streaming platforms.
    \item Case study: While we are working with some research labs in evaluating the FEVA for their video annotation \cite{Shrestha2022-jq}, we want to invite other interested research labs to try out FEVA, collaborate with us, and grow as a community to address needs that may not have been realized by our research so far.
\end{itemize}

\section{Conclusion}
We present a new event video annotation tool with streamlined interaction techniques and a dynamic UI, contextually visible, and active features organized based on the workflow and usage frequency. With features like speed labeling, users can accurately annotate videos in real-time. With simplified onboarding and workflow, researchers can set up and start annotating videos using minimal time. We release FEVA's source code in GitHub for everyone to try and further extend its features. The community can also find project samples, tutorial videos, GitHub issues for support, and future updates on the GitHub page. As we expand our case studies, we invite more researchers to use FEVA or contact us if you wish to collaborate.

\section*{Acknowledgments}
We thank Chethan Parameshwara, Levi Burner, Lindsay Little, and peers from UMD and the Perception and Robotics Group for their valuable feedback and discussions. We extend special thanks to all our project contributors Johnny Chiu, Rachelle Sims, John Gao, Leya Abraham, Vikram Sehgal, Swagata Chakroborty, Lucas Stuart, and Lin Chen. The support of NSF under grant OISE 2020624 is greatly acknowledged.

\bibliographystyle{unsrt}  
\bibliography{FEVA}

\clearpage
\appendix

\section{Online Resources}
Please visit FEVA website \url{http://www.snehesh.com/feva} for more information, code, instruction videos, samples, and any updates.

\section{List of Video Annotation Tools surveyed}
\label{appendix: list_of_VAT}
\begin{itemize}
    \item A Formative Study for Record-time Manual Annotation of First-person Videos
    \item A multi-level video annotation tool based on XML-dictionaries
    \item A Semi-Automatic Video Annotation Tool to Generate Ground Truth for Intelligent Video Surveillance Systems
    \item Advene
    \item AIBU
    \item An innovative web-based collaborative platform for video annotation
    \item An Ontology Web Application-based Annotation Tool for Intangible Culture Heritage Dance Videos
    \item Anvil
    \item atlas.ti
    \item Augmented Studio: Projection Mapping on Moving Body for Physiotherapy Education
    \item Automatic tagging of video based on voice and localization
    \item Automatically Freezing Live Video for Annotation during Remote Collaboration
    \item AVISA: An annotation tool for video understanding
    \item BeaverDam: Video Annotation Tool for Computer Vision Training Labels
    \item BEDA: Visual Analysis of Relationships between Behavioral and Physiological Sensor Data
    \item CASAM: collaborative human-machine annotation of multimedia
    \item CLIPPER: Audiovisual Annotation in the Study of Physics
    \item CoAT: A Web-based, Collaborative Annotation Tool
    \item CoVidA: Pen-based collaborative video annotation
    \item Crowd-Guided Ensembles: How Can We Choreograph Crowd Workers for Video Segmentation?
    \item CrowdSport: Crowd-based Semantic Event Detection and Video Annotation for Sports Videos
    \item DarkLabel
    \item Demo: Semantic Human Activity Annotation Tool - Using Skeletonized Surveillance Videos
    \item EagleView: A Video Analysis Tool for Visualising and Querying Spatial Interactions of People and Devices
    \item Elan
    \item Generating annotations for how-to videos using crowdsourcing
    \item Glance: rapidly coding behavioral video with the crowd
    \item HistoryTracker: Minimizing Human Interactions in Baseball Game Annotation
    \item iSeg: Semi-automatic ground truth annotation in videos: An interactive tool for polygon-based object annotation and segmentation
    \item iVAT: An interactive tool for manual, semi-automatic and automatic video annotation
    \item LabelMe
    \item Marquee: a tool for real-time video logging
    \item MediaDiver: viewing and annotating multi-view video
    \item MoViA: a mobile video annotation tool
    \item MRAS: Annotations for streaming video on the web
    \item Multimodal Video Annotation for Contemporary Dance Creation 
    \item MuLVAT: A Video Annotation Tool Based on XML-Dictionaries and Shot Clustering
    \item NVivo
    \item Oudjat is dedicated to the manual annotation facial expressions of emotion(FEE)
    \item Redesigning video analysis: an interactive ink annotation tool
    \item Rethinking Engagement with Online News through Social and Visual Co-Annotation
    \item Sirio, orione and pan: an integrated web system for ontology-based video search and annotation
    \item Stabilized Annotations for Mobile Remote Assistance
    \item SVAT
    \item The MESH mobile video annotation tool
    \item Timelinely
    \item Tool Eval: Rapid Model-Driven Annotation and Evaluation for Object Detection in Videos
    \item ToolScape: Enhancing the Learning Experience of How-to Videos
    \item VACA: a tool for qualitative video analysis
    \item VAnnotator: Annotations as multiple perspectives of video content
    \item VATIC
    \item VCode and VData: Illustrating a new Framework for Supporting the Video Annotation Workflow
    \item VIA: The VIA Annotation Software for Images, Audio and Video
    \item ViBAT
    \item VideoAnt
    \item VideoJot: A Multifunctional Video Annotation Tool
    \item VidOR: Annotating Objects and Relations in User-Generated Videos
    \item ViTBAT: Video Tracking and Behavior Annotation Tool
    \item VoTT
\end{itemize}

\clearpage
\section{FEVA Client Architecture}
\begin{figure}[ht]
  \label{fig: feva_architecture_client}
  \includegraphics[width=\textwidth]{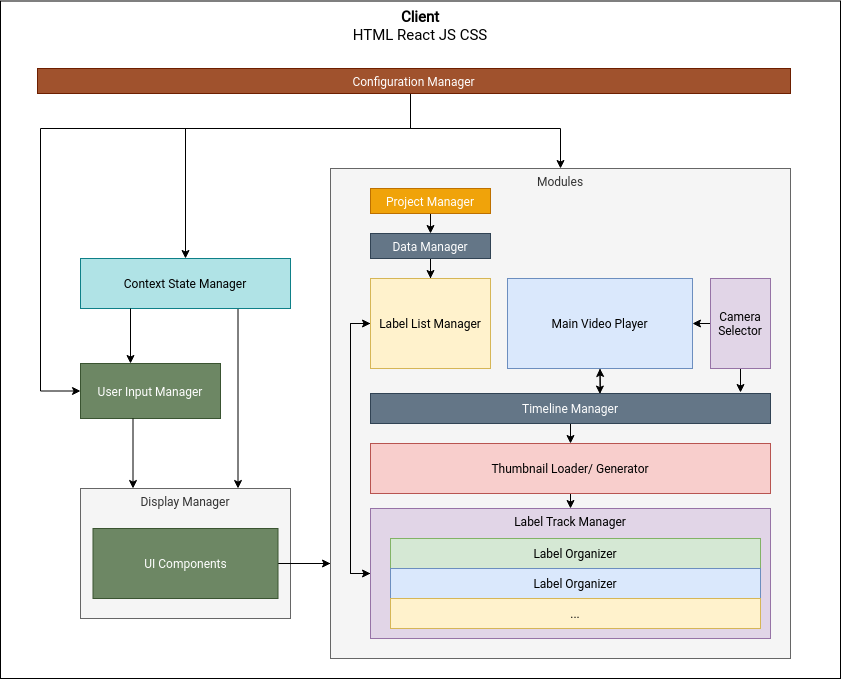}
  \caption{FEVA Client side block diagram of the various modules}
  \label{appendix: feva_client_architecture}
\end{figure}

\end{document}